\documentclass[conference]{IEEEtran}
\IEEEoverridecommandlockouts
\usepackage{cite}
\usepackage{amsmath,amsfonts,amssymb}
\usepackage{algorithmic}
\usepackage{graphicx}
\usepackage{textcomp}
\usepackage{xcolor}
\usepackage{makecell}
\usepackage[colorlinks = true,
            linkcolor = black,
            urlcolor  = blue,
            citecolor = black,
            anchorcolor = black]{hyperref}

\def\BibTeX{{\rm B\kern-.05em{\sc i\kern-.025em b}\kern-.08em
    T\kern-.1667em\lower.7ex\hbox{E}\kern-.125emX}}
\begin{document}

\title{Generative Iris Prior Embedded Transformer for 

Iris Restoration\\

\thanks{* Corresponding author.}
}

\author{
\IEEEauthorblockN{
Yubo Huang\IEEEauthorrefmark{2},
Jia Wang\IEEEauthorrefmark{3},
Peipei Li\IEEEauthorrefmark{3},
Liuyu Xiang\IEEEauthorrefmark{3},
Peigang Li\IEEEauthorrefmark{2},
Zhaofeng He*\IEEEauthorrefmark{3}}
\IEEEauthorblockA{\IEEEauthorrefmark{2}School of Integrated Circuits, Beijing University of Posts and Telecommunications, Beijing, China}
\IEEEauthorblockA{\IEEEauthorrefmark{3}School of Artificial Intelligence, Beijing University of Posts and Telecommunications, Beijing, China}
\IEEEauthorblockA{\{huangyubo, wangj, lipeipei, xiangly, pgli, zhaofenghe\}@bupt.edu.cn}
}

\maketitle

\begin{abstract}
Iris restoration from complexly degraded iris images, aiming to improve iris recognition performance, is a challenging problem. Due to the complex degradation, directly training a convolutional neural network (CNN) without prior cannot yield satisfactory results. In this work, we propose a \textbf{g}enerative iris prior embedded Trans\textbf{former} model (Gformer), in which we build a hierarchical encoder-decoder network employing Transformer block and generative iris prior. First, we tame Transformer blocks to model long-range dependencies in target images. Second, we pretrain an iris generative adversarial network (GAN) to obtain the rich iris prior, and incorporate it into the iris restoration process with our iris feature modulator. Our experiments demonstrate that the proposed Gformer outperforms state-of-the-art methods. Besides, iris recognition performance has been significantly improved after applying Gformer. Our code is available \href{https://github.com/sawyercharlton/Generative-Iris-Prior-Embedded-Transformer-for-Iris-Restoration}{here}.
\end{abstract}


\section{Introduction}
Iris recognition is a non-contact, high-security, and high-reliability identity authentication technology \cite{daugman2009iris}. It has been widely used in public security, financial banking, mobile Internet, and other fields. However, current iris recognition system still needs to be improved for supporting large-scale applications. One of the bottlenecks is the difficulty of acquiring high-quality iris images: (1) At long distances, the resolution of the acquired iris image is too low. (2) If a person stands outside the camera's depth of field, it creates an out-of-focus blur. (3) If a person is moving, motion blur occurs.
Therefore, iris restoration technology is significant to iris recognition. 

Traditional iris image super-resolution methods \cite{luk1993advanced, nguyen2011quality, hollingsworth2009iris} are often based on multi-view images or videos for high-quality image generation. With the development of deep learning \cite{lim2017enhanced, zhang2018image}, breakthroughs have been made in the field of image restoration, as well as iris restoration. Deep learning based iris restoration methods \cite{ribeiro2019iris, ribeiro2017exploring} often use a low-quality image as input to predict high-quality images, while use real high-quality images as supervised information to train the model. Among these convolutional neural network (CNN) models, the encoder-decoder-based U-shaped network \cite{ronneberger2015u} is commonly adopted since this architecture can represent images at multiple scales and has high computational efficiency. Similarly, we employ the U-shaped architecture in this work.
 
However, CNN cannot model long-range pixel dependencies, and the weight of the convolution filter cannot be flexibly adjusted according to the input content. To eliminate these issues, we replace CNN with a more powerful, dynamically working self-attention (SA), which calculates the response of a given pixel by weighted sums of other locations. Transformer \cite{vaswani2017attention, dosovitskiy2020image} excels in natural language processing and high-level visual tasks because SA can efficiently capture long-range pixel interactions. However, its computational complexity grows quadratically as the image resolution increases, making it infeasible to apply to high-resolution image restoration tasks. Referring to Mobilenets \cite{howard2017mobilenets}, we adopt a multi-head SA operating on the channel dimension, which has a linear computational complexity. This makes the application of Transformer in iris image restoration tasks more practical. 

Nevertheless, iris image has various microstructures, such as spots, stripes, filaments, and crypts \cite{daugman2001epigenetic}. These microstructures and the background constitute the rich texture of the iris modality \cite{sun2008ordinal}. With iris prior, CNN-based methods or Transformer-based methods are easier to precisely restore these complicated iris microstructures. Several methods  \cite{abdal2019image2stylegan, karras2017progressive} use pretrained generative adversarial network (GAN)  \cite{goodfellow2020generative} models  as prior, which map the input image to the closest latent code, and generate the corresponding output. However, these methods only contain the latent codes of low-resolution images and provide insufficient guidance for restoration work. In this paper, we leverage features of multi-resolution images to guide iris restoration with our hierarchical architecture. We pretrain an iris GAN as a generative iris prior and incorporate features of multi-resolution images using our iris feature modulator. Generative iris prior can map the feature extracted by Transformer to its closest latent code.

The contributions of our method are as follows:
(1) We propose an iris restoration method called Gformer, a U-shaped network with an encoder-decoder structure. Our iris feature modulator connects the encoder and the decoder. (2) The encoder employs Transformer blocks, while the decoder encapsulates generative iris prior we pretrained. The iris feature modulator embeds iris prior into Transformer. (3) Extensive experiments show that our Gformer outperforms state-of-the-art methods for restoring iris images that suffered complex degradation and improving recognition performance.

\section{Proposed Method}
\begin{figure*}[t]
% \centerline{\epsfig{figure=image1.ps,width=8.5cm}}
\includegraphics[width=18cm]{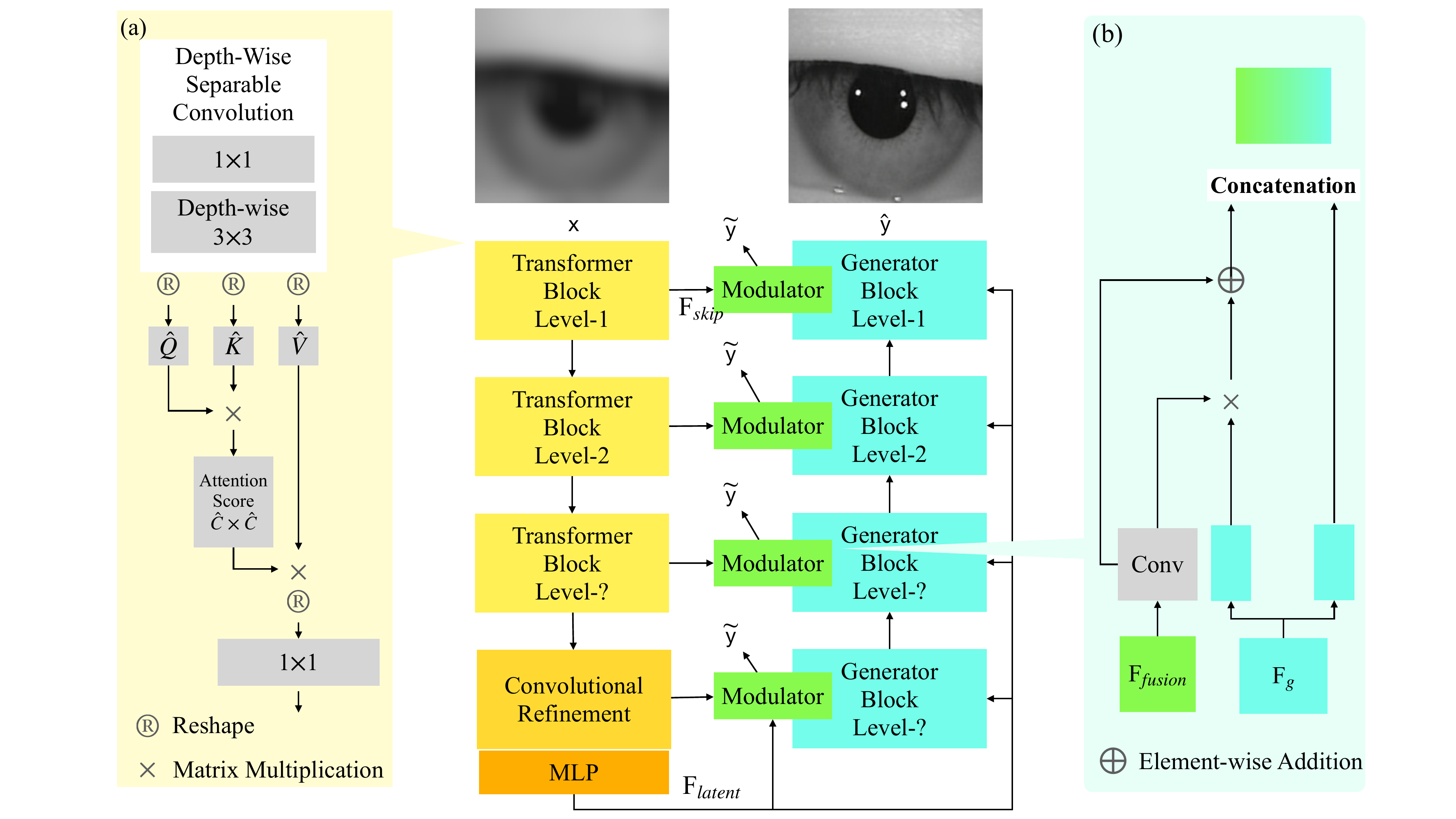}
\caption{Overview of Gformer architecture. Gformer consists of Transformer encoder and generative iris prior embedded decoder. They are bridged by the iris feature modulator. (\textbf{a}) shows the core component of Transformer block: depth-wise self-attention that operates \textit{attention score} across channels rather than spatial dimension. (\textbf{b}) shows the spatial feature transform in the iris feature modulator, which leverages generative iris prior.}
\label{fig:arc}
\end{figure*}

\subsection{Overview}
Our Gformer follows a U-shaped encoder-decoder architecture. We take a low-quality iris image undergoing complex degradation as the input.

The encoder consists of multi-scale Transformer blocks, convolutional refinement layers, and fully connected layers. We adopt depth-wise self-attention and depth-wise feed-forward networks in Transformer blocks. The Transformer block on each level will output a feature of corresponding size as a skip feature map $\textbf{\textit{F}}_{skip}$ connecting the decoder.

The number of encoder blocks must equal the number of generator blocks to make the model run successfully. However, too many Transformer blocks lead to too much memory overhead when training. We reduce three Transformer blocks and add three convolution blocks to match the generator blocks. In this paper,  $\#Transformer\_blocks=4, \#Convolution\_blocks=3, \#generator\_blocks=7$. 

The multi-layer perceptron (MLP) at the end of the encoder outputs a deep-level feature called a latent code $\textbf{\textit{F}}_{latent}$, which is broadcast to multiple scales and fed into the decoder. MLP can effectively preserve semantic property, which is helpful for high-level performance, as previous work \cite{zhu2020domain} has demonstrated.

We first train StyleGANv2 \cite{karras2020analyzing} with iris images as prior and use the generator of StyleGANv2 as the backbone of the decoder. The decoder also consists of multi-scale generator blocks, the number of which is equal to the number of skip feature maps $\textbf{\textit{F}}_{skip}$. 

We design modulators attached to every generator block. Our modulators 1) modulate skip feature maps $\textbf{\textit{F}}_{skip}$ and latent codes $\textbf{\textit{F}}_{latent}$, 2) generate pyramid iris images before leveraging prior, which are later used to calculate pyramid loss, and 3) modulate generative iris prior using channel-split spatial feature transform.

\subsection{Depth-Wise Self-Attention}

The computational overhead of Transformers comes mainly from self-attention. In the conventional self-attention \cite{vaswani2017attention, dosovitskiy2020image}, \textit{attention score} is the key(\textbf{K})-query(\textbf{Q}) dot-product interaction: $\alpha=\textbf{Q}\cdot\textbf{K}$. The time and memory complexity is $\mathcal{O}(\textit{W}^{2}\textit{H}^{2})$, $\textit{W} \times \textit{H}$ is the resolution of an image. The computational complexity of \textit{attention score} increases quadratically with the image resolution. Therefore, such self-attention in image restoration tasks will incur excessive computational overhead. To reduce the computational cost, we employ depth-wise separable convolution (DSC) \cite{howard2017mobilenets} in self-attention, shown in Fig.~\ref{fig:arc}(a). In DSC, the first step is to use pixel convolution, a 1$\times$1 convolution that collects content information about a pixel across all channels; The second step is to use depth-wise convolution to extract content information from spatial dimension in a channel. From a layer normalized tensor $\textbf{Y}\in\mathbb{R}^{\hat{H}\times\hat{W}\times\hat{C}}$, DSC generates \textit{query} (\textbf{Q}), \textit{query} (\textbf{K}), \textit{query} (\textbf{V}). Before calculating \textit{attention score} $\beta$, we reshape \textbf{Q}, \textbf{K}, \textbf{V} to make our \textit{attention score} $\beta\in\mathbb{R}^{\hat{C}\times\hat{C}}$ because the size of traditional \textit{attention score} $\alpha\in\mathbb{R}^{\hat{H}\hat{W}\times\hat{H}\hat{W}}$ \cite{vaswani2017attention, dosovitskiy2020image} is too huge. After this, $\hat{\textbf{Q}}\in\mathbb{R}^{\hat{H}\hat{W}\times\hat{C}}$, $\hat{\textbf{K}}\in\mathbb{R}^{\hat{H}\hat{W}\times\hat{C}}$, and $\hat{\textbf{V}}\in\mathbb{R}^{\hat{H}\hat{W}\times\hat{C}}$ are generated. Our \textit{attention score} is formulated as:

\begin{equation}
\beta=\hat{\textbf{Q}}\cdot\hat{\textbf{K}}
\end{equation}
Our depth-wise self-attention is formulated as:
\begin{equation}
Attention\left(\hat{\textbf{Q}},\hat{\textbf{K}},\hat{\textbf{V}}\right)=\hat{\textbf{V}} \cdot Softmax \left( \theta \times \beta \right),
\end{equation}
where $\theta$ is a learnable scaling parameter.

Similar to DSA, we use depth-wise convolutions in the feed-forward network. Depth-wise convolution encodes information from neighboring pixels, which can effectively learn the local image structure for restoration. Overall, the depth-wise feed-forward network enriches the iris texture by cooperating with the depth-wise self-attention.

\subsection{Iris Feature Modulator}

To embed generative iris prior, we design an iris feature modulator, which leverages skip feature map $\textbf{\textit{F}}_{skip}$ and latent code $\textbf{\textit{F}}_{latent}$ output of the decoder to modulate the features of the StyleGANv2 generator block output, shown in Fig.~\ref{fig:arc}. The first step is to fuse the skip feature map $\textbf{\textit{F}}_{skip}$ and the latent code $\textbf{\textit{F}}_{latent}$,
\begin{equation}
    \begin{aligned}
        \textbf{\textit{F}}_{fusion}^{(i)}&=Conv_{fusion}\left(\textbf{\textit{F}}_{fusion}^{(i-1)}+\textbf{\textit{F}}_{skip}^{(i)}\right) \\
        \textbf{\textit{F}}_{fusion}^{(0)}&=\textbf{\textit{F}}_{latent},
    \end{aligned}
\end{equation}
where $\textbf{\textit{F}}_{fusion}^{(i)}$ and $\textbf{\textit{F}}_{skip}^{(i)}$ represent the fused feature and skip feature map in i-th block. $\textbf{\textit{F}}_{fusion}$ and $\textbf{\textit{F}}_{latent}$ are fed into the generator block to get $\textbf{\textit{F}}_{g}$:

\begin{equation}
    \textbf{\textit{F}}_{g}=Generator\left(\textbf{\textit{F}}_{fusion}, \textbf{\textit{F}}_{latent}\right)
\end{equation}

Our iris feature modulator will further modulate $\textbf{\textit{F}}_{g}$ using channel-split spatial feature transform(CS-SFT) \cite{wang2018recovering, wang2021towards}: divide it into two parts equally in the channel dimension, one part remains unchanged, the other part is modulated, and finally concatenate the two parts to obtain the output feature:
\begin{equation}
    \begin{aligned}
        \textbf{\textit{F}}_{output}&=Concat\left(\textbf{\textit{F}}_{g}^{split0}, \mu\odot\textbf{\textit{F}}_{g}^{split1}+\sigma\right) \\
        \mu, \sigma&=Conv\left(\textbf{\textit{F}}_{fusion}\right),
    \end{aligned}
\end{equation}
where $\textbf{\textit{F}}_{g}^{split0}$ and $\textbf{\textit{F}}_{g}^{split1}$ are split features from $\textbf{\textit{F}}_{g}$ in channel dimension, $\mu$ and $\sigma$ are scaling and shifting factors. Our iris feature modulator can not only preserve the identity information of the iris, but also generate high-quality iris texture. Additionally, its computational complexity is smaller than methods modulating the whole channels.

\subsection{Model Objectives}

To restore low-quality iris images and achieve higher recognition performance, we adopt four loss functions: L1 loss $\textit{L}_{l1}$, perceptual loss $\textit{L}_{per}$, adversarial loss $\textit{L}_{adv}$, and pyramid loss $\textit{L}_{pyr}$:
\begin{equation}
    \textit{L}_{l1}=\|\textbf{y}-\hat{\textbf{y}}\|_{1}
\end{equation}
where $\textbf{y}$ denotes original high-quality iris images, and $\hat{\textbf{y}}$ denotes restored iris images. L1 loss $\textit{L}_{l1}$ constraints the outputs $\hat{\textbf{y}}$ close to the ground-truth $\textbf{y}$.
\begin{equation}
    \textit{L}_{per}=\|\phi\left(\textbf{y}\right)-\phi\left(\hat{\textbf{y}}\right)\|_{1},
\end{equation}
where $\phi$ is the pretrained VGG-19 network  \cite{simonyan2014very}. Perceptual loss $\textit{L}_{per}$ further enhances the iris texture.
\begin{equation}
    \textit{L}_{adv}=\textit{E}_{\hat{\textbf{y}}}\log\left(1+\exp\left(-\textit{D}\left(\hat{\textbf{y}}\right)\right)\right),
\end{equation}
where \textit{D} denotes the discriminator. The adversarial loss $\textit{L}_{adv}$ is used for recovering more realistic iris images with vivid details.
\begin{equation}
    \textit{L}_{pyr}=\|\textbf{y}-\widetilde{\textbf{y}}\|_{1},
\end{equation}
we convert the fusion features $\textbf{\textit{F}}_{fusion}^{(i)}$ of each level into images of the corresponding resolution, denoted as $\widetilde{\textbf{y}}$. $\widetilde{\textbf{y}}$ is the intermediate RGB output. To produce it, just feed $\textbf{\textit{F}}_{fusion}^{(i)}$ to a convolutional layer $(output\_channels=3, kernel\_size=1)$. Then, we use L1 loss to constraint the restored iris images of different sizes closer to the ground-truth before embedding prior. Arranged together from small to large, these images look like pyramids, so this loss function is called pyramid loss. The final loss \textit{L} is as follows:
\begin{equation} \textit{L}=\theta_{1}\textit{L}_{l1}+\theta_{2}\textit{L}_{per}+\theta_{3}\textit{L}_{adv}+\textit{L}_{pyr},
\end{equation}
where $\theta_{1}$, $\theta_{2}$ and $\theta_{3}$ are balancing parameters. We empirically set $\theta_{1}=0.1$, $\theta_{2}=1$ and $\theta_{3}=0.1$.

\section{Experiments}

\begin{figure*}
\centering
\begin{minipage}[t]{0.11\textwidth}
\centering
\includegraphics[width=1\linewidth]{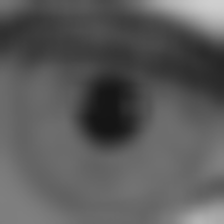}
\includegraphics[width=1\linewidth]{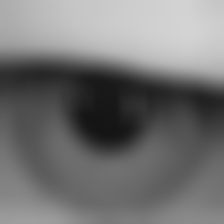}
\centerline{Input}\medskip
\end{minipage}
\begin{minipage}[t]{0.11\textwidth}
\centering
\includegraphics[width=1\linewidth]{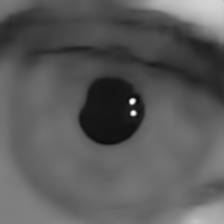}
\includegraphics[width=1\linewidth]{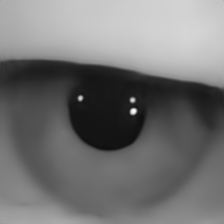}
\centerline{EDSR}\medskip
\end{minipage}
\begin{minipage}[t]{0.11\textwidth}
\centering
\includegraphics[width=1\linewidth]{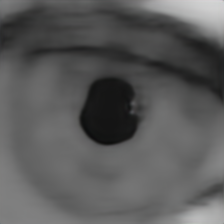}
\includegraphics[width=1\linewidth]{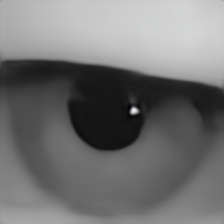}
\centerline{RCAN}\medskip
\end{minipage}
\begin{minipage}[t]{0.11\textwidth}
\centering
\includegraphics[width=1\linewidth]{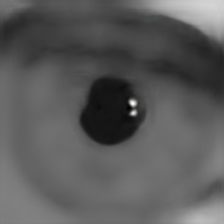}
\includegraphics[width=1\linewidth]{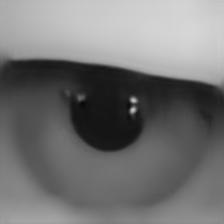}
\centerline{SRGAN}\medskip
\end{minipage}
\begin{minipage}[t]{0.11\textwidth}
\centering
\includegraphics[width=1\linewidth]{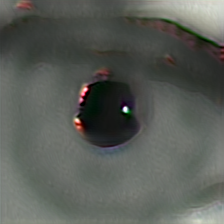}
\includegraphics[width=1\linewidth]{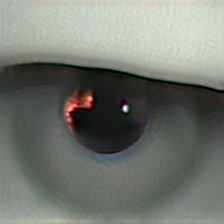}
\centerline{ESRGAN}\medskip
\end{minipage}
\begin{minipage}[t]{0.11\textwidth}
\centering
\includegraphics[width=1\linewidth]{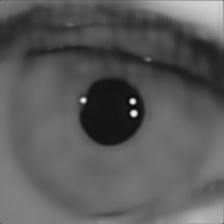}
\includegraphics[width=1\linewidth]{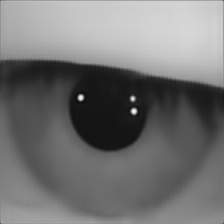}
\centerline{DeblurGANv2}\medskip
\end{minipage}
\begin{minipage}[t]{0.11\textwidth}
\centering
\includegraphics[width=1\linewidth]{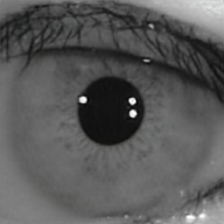}
\includegraphics[width=1\linewidth]{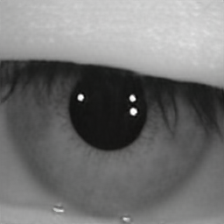}
\centerline{\textbf{Gformer}}\medskip
\end{minipage}
\begin{minipage}[t]{0.11\textwidth}
\centering
\includegraphics[width=1\linewidth]{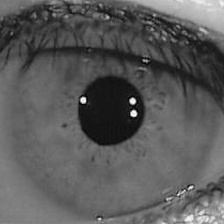}
\includegraphics[width=1\linewidth]{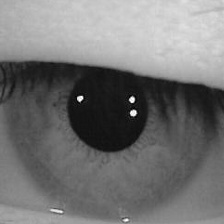}
\centerline{Ground-Truth}\medskip
\end{minipage}
\caption{Qualitative comparison with state-of-the-art methods.}
\label{fig:qua}
\end{figure*}

\begin{table*}[t]
\begin{center}
\caption{Quantitative comparison with state-of-the-art methods.} \label{tab:com}
\begin{tabular}{ccc ccc ccc}
  \hline
  % after \\: \hline or \cline{col1-col2} \cline{col3-col4} ...
Method & AUC $\uparrow$ & ERR $\downarrow$	& \makecell{TAR@FAR \\ =0.001 $\uparrow$} & \makecell{TAR@FAR \\ =0.01 $\uparrow$} & \makecell{TAR@FAR \\ =0.1 $\uparrow$} & PSNR $\uparrow$ & SSIM $\uparrow$ & \makecell{Inference \\ Time $\downarrow$} \\ \hline
Input & 0.9521 & 0.1153 & 0.1992 & 0.5361 & 0.8658 & 30.68 & 0.8336 & - \\
EDSR & 0.9796 & 0.0705 & 0.5356 & 0.7538 & 0.9496 &	\textbf{34.15} & \underline{0.8808} & 88ms \\
RCAN & 0.9818 & 0.0679 & 0.5608 & 0.7666 & 0.9558 & \underline{33.57} & 0.8789 & 70ms \\
SRGAN & 0.9732 & 0.0853	& 0.3960 & 0.6705 & 0.9279 & 32.85 & 0.8695 & \textbf{10ms} \\
ESRGAN & 0.9585	& 0.1093 & 0.4816 & 0.6876 & 0.8839 & 30.29 & 0.8314 &  154ms \\
DeblurGANv2 & \underline{0.9890} & \underline{0.0492}& 0.6995 & 0.8659 & \textbf{0.9753} & 32.86 & 0.8785 & \underline{15ms} \\
HiFaceGAN & 0.9869 & 0.0529 & \underline{0.7549} & \underline{0.8843} & 0.9638 & 27.74 & 0.8172 & 91ms \\
SISN & 0.9836 & 0.0542 & 0.5878 & 0.8274 & 0.9679 & 32.36 & 0.8729  & 202ms \\
SparNet & 0.9871 & 0.0553 & 0.6386 & 0.8357 & 0.9709 & 31.94 & 0.8660 & 134ms \\
\textbf{Gformer (ours)} & \textbf{0.9898} & \textbf{0.0455} & \textbf{0.8074} & \textbf{0.9068} & \underline{0.9725} & 33.48 & \textbf{0.8815} & 146ms \\ \hline
Ground-Truth & 0.9914 & 0.0274 & 0.9323& 0.9613 & 0.9833 & - & -  & - \\ \hline
\end{tabular}
\end{center}
\end{table*}

\begin{figure}[t]
% \centerline{\epsfig{figure=image1.ps,width=8.5cm}}
\includegraphics[width=8.5cm]{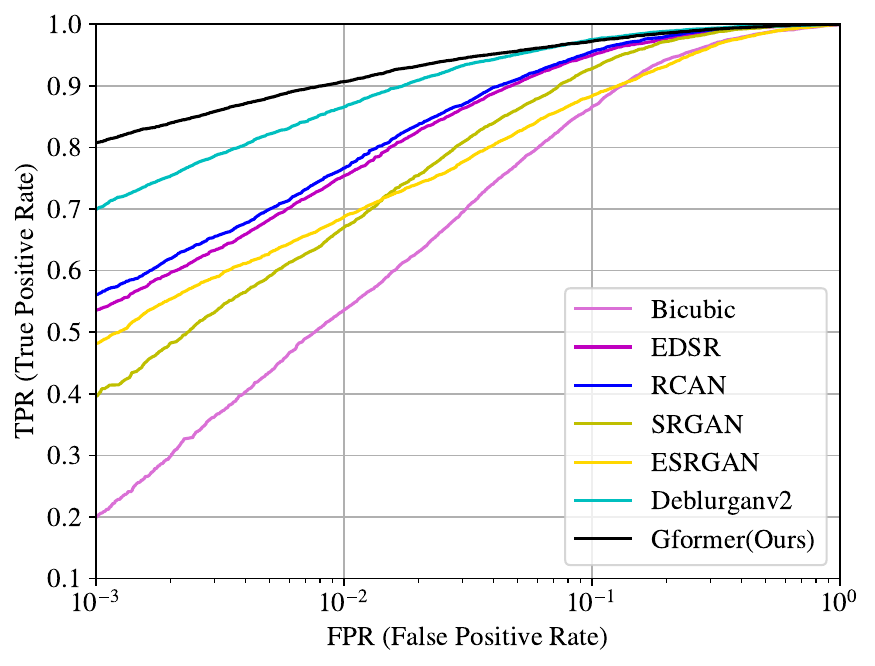}
\caption{ROC curve for comparison with state-of-the-art methods.}
\label{fig:roc}
\end{figure}

\subsection{Datasets}\label{Datasets}
In order to verify the feasibility of our Gformer, we conduct experiments on the large-scale datasets CASIA-Iris-Lamp and CASIA-Iris-Thousand. To advance research on large-scale iris recognition, Institute of Automation, Chinese Academy of Sciences (CASIA) constructed the CASIA-Iris V4 dataset, which includes CASIA-Iris-Lamp and CASIA-Iris-Thousand. CASIA-Iris-Lamp contains 822 eye images of 411 subjects, including 822 categories with an image resolution of $640\times480$. CASIA-Iris-Thousand contains 20,000 eye images of 640$\times$480 from the eyes of 1,000 subjects, including 2,000 categories. We combine, screen, position, and crop them, resulting in 28,089 iris images, including 2,738 categories, with an image resolution of $224^2$.

During training, we enlarge the image resolution to $256^2$ as input. We use synthetic data for training. To make the training dataset as close as possible to the real-world low-quality iris images, we propose the following degradation model: 

\begin{equation}            
\textbf{x}=\left(\textbf{y}\circledast\textbf{g}_{\delta}\circledast\textbf{m}_{\gamma}\right)\downarrow_{\tau} .
\end{equation}

The high quality image \textbf{y} is convolved with Gaussian blur kernel $\textbf{g}_{\delta}$ and motion blur kernel $\textbf{g}_{\gamma}$, and downsampled with a scaling factor $\tau$. For each training pair, we randomly sample $\delta$, $\gamma$, and $\tau$ from [3:15], [3:15], and [1:4], respectively.

The testing dataset is synthesized in the same way as the training dataset. Additionally, there is no overlap between the testing dataset and the training dataset.

\subsection{Implementation Details}

The generative iris prior we used is a pretrained StyleGANv2 with an output of $256^2$. The channel multiplication factor is set to 1 (the default is 2 \cite{karras2020analyzing}), and other parameters follow the default settings.

In order to accommodate the channel number of StyleGANv2 latent code 512 and the input image resolution $256^2$, our Gformer employs a 4-level Transformer encoder. From level-1 to level-4, the number of attention heads are $[1, 2, 4, 8]$, the number of channels are $[64,128,256,512]$, and the output image resolutions are $[256^2, 128^2, 64^2, 32^2]$. Each level contains 1 Transformer block. The channel multiplication coefficient $\epsilon$ is set to $\epsilon=2.66$ referring to Restormer \cite{zamir2022restormer}. The convolutional refinement module is a 3-level hierarchical network following Transformer blocks, the number of channels is all 512, and the output image resolutions are $[16^2, 8^2, 4^2]$.

During training, we adopt the Adam optimizer with a batch size of 16. The learning rate is set to $2\times10^{-4}$. We implement our model with the PyTorch framework, and train it using NVIDIA Tesla v100 GPUs.

\subsection{Comparisons with State-of-the-Art Methods}

\begin{table*}[t]
\begin{center}
\caption{Ablation study results.} \label{tab:abl}
\begin{tabular}{ccc ccc ccc}
  \hline
  % after \\: \hline or \cline{col1-col2} \cline{col3-col4} ...
Method & AUC $\uparrow$ & ERR $\downarrow$	& \makecell{TAR@FAR \\ =0.001 $\uparrow$} & \makecell{TAR@FAR \\ =0.01 $\uparrow$} & \makecell{TAR@FAR \\ =0.1 $\uparrow$} & PSNR $\uparrow$ & SSIM $\uparrow$ & LPIPS $\downarrow$ \\ \hline
w/o Generative Iris Prior & 0.9864 & 0.0532 & 0.7475 & 0.8773 &  0.9655 & 32.82 & 0.8754 & 0.1513 \\
w/o Transformer & 0.9883 & 0.0505 & 0.7843 & 0.8873 & 0.9689 & 32.88 & 0.8631 & 0.1472 \\
w/o Spatial Feature Transform & 0.9879 & 0.0480 & 0.7721 & 0.8917 & 0.9736 & 32.12 & 0.8582 & 0.1586 \\
w/o Chanel-Split & \underline{0.9903} & \textbf{0.0435} & 0.7889 & \underline{0.9060} & \textbf{0.9765} & 32.96 & 0.8734 & 0.1466 \\
w/o Pyramid Loss & \textbf{0.9905} & 0.0456 & \underline{0.7908}  & 0.9033 & \underline{0.9751} & \underline{33.42} & \underline{0.8791} & \underline{0.1411} \\
w/o Perceptual Loss & 0.9521 & 0.1252 & 0.5266 & 0.6725 & 0.8572 & 29.50 & 0.8068 & 0.2056 \\
w/o Adversarial Loss & 0.9864 & 0.0532 & 0.7200 & 0.8702 & 0.9665 & 32.12 & 0.8629 & 0.1565 \\
Overall & 0.9898 & \underline{0.0455} & \textbf{0.8074} & \textbf{0.9068} & 0.9725 & \textbf{33.48} & \textbf{0.8815} & \textbf{0.1396} \\
  \hline
\end{tabular}
\end{center}
\end{table*}

We compare our Gformer with several state-of-the-art image restoration methods: EDSR  \cite{lim2017enhanced}, RCAN  \cite{zhang2018image}, SRGAN  \cite{ledig2017photo}, ESRGAN  \cite{wang2018esrgan}, DeblurGANv2  \cite{kupyn2019deblurgan}, HiFaceGAN  \cite{Yang2020HiFaceGANFR}, SISN  \cite{lu2021face}, and SparNet  \cite{khoei2020sparnet}. For a fair comparison, we retrained these models on the dataset we mentioned in section \ref{Datasets}.

The goal of iris image restoration differs from that of general image restoration. Iris image restoration aims to improve the performance of iris recognition. Enhancing iris image quality and visual effects does not necessarily improve recognition performance \cite{alonso2018survey}. We adopt AUC, ERR, TAR@FAR=0.001, TAR@FAR=0.01, and TAR@FAR=0.1, as evaluation metrics. EER is the abbreviation of equal error rate, which is an extremely critical error rate metric in iris recognition. TAR indicates true acceptance rate, while FAR means false acceptance rate. The smaller FAR, the stricter the identification criteria. Therefore, the metric TAR@FAR=0.001 can best reflect the quality of the algorithm recognition performance. Firstly, we use the values of TAR@FAR=0.1, TAR@FAR=0.01, and TAR@FAR=0.001 to draw the receiver operating characteristic (ROC) curve. Then, we calculate the area under the curve (AUC), that is, the lower clamp area of the ROC curve. The larger the AUC, the better the recognition performance. We also adopt pixel-wise metrics (PSNR and SSIM) and the perceptual metric (LPIPS  \cite{zhang2018unreasonable}). Finally, we conduct an inference time comparison.

The qualitative comparisons are shown in Fig.~\ref{fig:qua}. Gformer produces plausible and realistic irises on complicated degradation, while other methods fail to recover faithful iris details. According to the ROC curve in Fig.~\ref{fig:roc} and the quantitative measures in Table ~\ref{tab:com}, it can be seen that our Gformer achieves the highest AUC, TAR@FAR=0.001, and TAR@FAR=0.01, and obtains lowest EER with a moderate level of inference time. However, Gformer achieves just ordinary PSNR and SSIM. This shows that the precision and realism of our method have unique advantages in iris image restoration tasks. Our Gformer can still achieve the best recognition performance improvement without improving image quality too well.

\subsection{Ablation Study}

To better understand the roles of components in Gformer and the training strategy, we conduct an ablation study by introducing some variants of Gformer. The evaluation scores are reported in Table~\ref{tab:abl}.

\textbf{Without generative iris prior}. Keeping the network structure unchanged, we do not use the pretrained weights of iris GAN. During training, we noticed a slowdown in the training speed. Besides, performance drops on all metrics.

\textbf{Without Transformer}. We use convolution block to replace Transformer block. During training, we find that the learning speed is slower, and we need to set the learning rate to 10 times the original. Performance degradation is severe on all metrics. 

\textbf{Without spatial feature transform}. Instead of using spatial feature transform (SFT) in the iris feature modulator, we simply add up $\textbf{\textit{F}}_{fusion}$ and $\textbf{\textit{F}}_{g}$. PSNR, SSIM, and LPIPS decline seriously. Besides, recognition performance EER and TAR@FAR=0.001 degrade to some extent. These show that SFT in the iris feature modulator is helpful for iris restoration, especially for image quality improvement.

\textbf{Without channel-split spatial feature transform}. Spatial feature transform is performed across all channels instead of half of the channels in Gformer. The computational complexity increases during training. AUC, EER, and TAR@FAR=0.001 grow, while others drop. This indicates that channel-split spatial feature transform can balance the complexity and performance of the model.

\textbf{Without pyramid loss}. AUC and TAR@FAR=0.001 rise, while the other metrics decline, notably the image quality metrics PSNR, SSIM. It shows that pyramid loss does not help or even impair recognition performance, but it helps a lot to enhance image quality. 

\textbf{Without perceptual loss}. All metrics drop dramatically, so perceptual loss is most significant to our method, which is very sensitive to the hyper-parameter of perceptual loss.

\textbf{Without adversarial loss}. All metrics decline to varying degrees. That means our method is sensitive to the hyper-parameter of adversarial loss.

\section{Conclusion}

In this paper, we proposed a generative iris prior embedded Transformer model for iris restoration, namely Gformer. Gformer can conduct multi-scale local-global representation learning on high-resolution images due to the hierarchical architecture. Specifically, our Transformer block consists of depth-wise self-attention and depth-wise feed-forward network, having linear complexity rather than quadratic. It can efficiently capture long-range pixel interactions for restoration. Besides, generative iris prior makes restoring complicated iris microstructures easier. The proposed iris feature modulator incorporates generative iris prior and features extracted by Transformer into the restoration process. Extensive experiments demonstrate that Gformer achieves superior performance on iris restoration and helps improve iris recognition.

\section*{Acknowledgment}

This work is supported by the National Natural Science Foundation of China under Grants 62176025, 62106015 and U21B2045.

\bibliographystyle{IEEEtran}
\bibliography{IEEEabrv,icme2022template}

\end{document}